\documentclass[9pt,twocolumn,twoside]{pnas-new}

\usepackage{bm}
\usepackage{xcolor}

\templatetype{pnasresearcharticle}

\title{Network structure of cascading neural systems predicts stimulus propagation and recovery}

\author[a]{Harang Ju}
\author[b]{Jason Z. Kim}
\author[b,c,d,e,f,g,1]{Danielle S. Bassett}

\affil[a]{Neuroscience Graduate Group, University of Pennsylvania, Philadelphia,
PA 19104 USA}
\affil[b]{Department of Bioengineering, University of Pennsylvania, Philadelphia, PA 19104 USA}
\affil[c]{Department of Electrical \& Systems Engineering, University of Pennsylvania, Philadelphia, PA 19104 USA}
\affil[d]{Department of Physics \& Astronomy, University of Pennsylvania, Philadelphia, PA 19104 USA}
\affil[e]{Department of Neurology, University of Pennsylvania, Philadelphia, PA 19104 USA}
\affil[f]{Department of Psychiatry, University of Pennsylvania, Philadelphia, PA 19104 USA}
\affil[g]{Santa Fe Institute, 1399 Hyde Park Rd, Santa Fe, NM 87501 USA}

\leadauthor{Ju}

\significancestatement{Cortical neurons become spontaneously active, producing bursts of electrical spikes. These neurons spike in well-characterized patterns of activity called cascades. However, these cascades have yet to be mechanistically linked to the patterns of connections among cortical neurons that underlie such activity. Here, we use theories of linear systems, network science, and information to demonstrate how the underlying patterns of connections influence the recovery of information through long-lasting cascades in empirical and simulated spiking neurons. Importantly, we show how one can use high-dimensional data of spontaneous neural spikes to reveal how the structure of networks can shape the processing of information, which can ultimately determine the computations that underlie cognition.}

\authorcontributions{
H.J., J.Z.K., and D.S.B. designed research;
H.J. and J.Z.K. performed research;
H.J. and J.Z.K. contributed analytic tools;
H.J. analyzed data;
H.J., J.Z.K., and D.S.B. wrote the paper
}
\authordeclaration{The authors declare no competing interest.}
\correspondingauthor{\textsuperscript{1}To whom correspondence should be addressed. E-mail: dsb@upenn.edu}


\begin{abstract}
Many neural systems display cascading behavior characterized by uninterrupted sequences of neuronal firing. This gap precludes an understanding of how variations in network structure manifest in neural dynamics and either support or impinge upon information processing. Here, we develop a theoretical understanding of how network structure supports information processing through network dynamics, and we validate our theory with empirical data. Using a generalized spiking model and mathematical tools from linear systems theory, network control theory, and information theory, we show how network structure can be designed to temporally extend the propagation and recovery of certain stimulus patterns. Moreover, we observe cycles as structural and dynamic motifs that are prevalent in such networks. Broadly, our results demonstrate how cascading neural networks could contribute to cognitive faculties that require lasting activation of neuronal patterns, such as working memory or attention.
\end{abstract}

\dates{This manuscript was compiled on \today}
\doi{\url{www.pnas.org/cgi/doi/10.1073/pnas.XXXXXXXXXX}}

\begin{document}

\maketitle
\thispagestyle{firststyle}
\ifthenelse{\boolean{shortarticle}}{\ifthenelse{\boolean{singlecolumn}}{\abscontentformatted}{\abscontent}}{}

\dropcap{A} central question in neuroscience is how connections between neurons determine patterns of neurophysiological activity that support organism function. Networks of neurons receive incoming stimuli and perform computations to shape cognition and behavior, such as the visual recognition of faces in regulating social behavior \cite{adolphs2003}. While many studies laud the ultimate goal of determining how the brain's network structure supports information processing \cite{watts1998, honey2007}, it remains challenging to empirically study the direct interactions between neural dynamics, connectivity, and computation. Indeed, neural connections and their underlying computational function have often been inferred through neural dynamics, and formal studies probing mechanistic relations among the three components have remained largely theoretical \cite{hopfield1982, benyishai1995, wang2002}.

One characteristic empirical feature of many systems is cascading dynamics, in which neurons display spontaneous bursts of activity. While these bursts may seem arbitrary, they actually comprise stochastic cascades that follow spatiotemporal patterns of activity \cite{haldeman2005}. These cortical cascading dynamics have been well-characterized in the empirical literature using a range of methods \textit{in vitro} \cite{beggs2003, beggs2004}, \textit{in vivo} \cite{gireesh2008, petermann2009, hahn2010, shriki2013, bellay2015, poncealvarez2018}, and \textit{ex vivo} \cite{shew2015} in a variety of organisms, including humans. In a complementary line of theoretical work, these neural systems have been hypothesized to operate within a regime that maximizes information transmission \cite{beggs2003, shew2011}, information storage \cite{haldeman2005}, computational power \cite{bertschinger2004}, and dynamic range \cite{kinouchi2006, shew2009, larremore2011b}. However, often left implicit in these analyses is the structure of the networks underlying such dynamics and how the structure may constrain those dynamics. Relatively recent empirical data show evidence of specific patterns of cortical connectivity. Cortical neurons are often strongly, bidirectionally connected to each other \cite{wang2006, lefort2009, ko2011}, potentially supporting optimal information storage \cite{brunel2016}. They also form higher-order network motifs in clusters of neurons \cite{markram1997, song2005, perin2011}, which in turn group into communities of neurons that perform most of the computation performed in a network \cite{shimono2014, nigam2016, faber2019}. These features of network structure have yet to be linked to the dynamics and computation that is  supported by those same circuits.

Here, we address this gap in knowledge through a series of analyses on simulations and empirical data. We first frame spike propagation as state transitions in a Markov chain to show that network structure constrains system memory through sustained activity. We then apply linear systems theory to predict distributions of cascade duration in the stochastic dynamics of simulated and empirical spiking neural networks. We find that cycles and strong connections in cycles---both of which are empirically observed network motifs---are notable contributors to the long tails in distributions of cascade duration. Finally, we use mutual information to probe the relations among network structure, cascade duration, and the information maintained in a network in 4 commonly studied generative graph models. Moreover, our method can accomodate networks that are both non-critical \cite{touboul2010, friedman2012, priesemann2014, touboul2017} and critical, and that show avalanche behavior, characterized by power-law distributions of cascade \textit{size} (i.e., the number of neurons that spike in a cascade) and \textit{duration}. Collectively, our findings show that the network topology reported extensively in the empirical literature can produce complex cascading dynamics through which a network can support the lasting activation of a cluster of neurons, which in turn allows for the discrimination of stimulus patterns implicated in working memory \cite{goldmanrakic1995, durstewitz2000, eriksson2015}.

\section*{Mathematical Framework}
\label{sec:mathframe}

\subsection*{Network formulation}
\begin{figure}[!ht]
\centering
\includegraphics[width=0.94\linewidth]{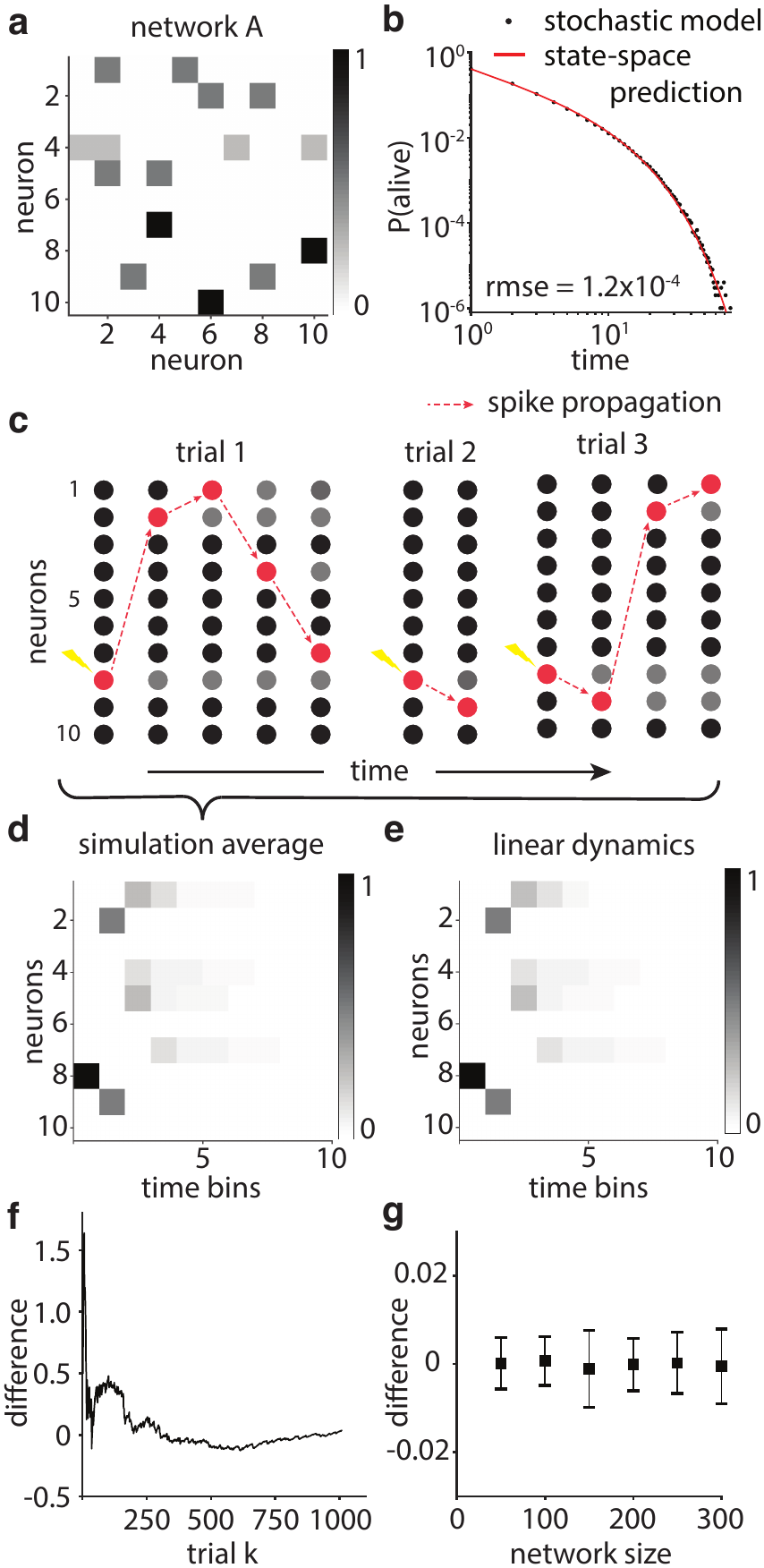}
\caption{
\textbf{A linear dynamical system accurately estimates the average spiking of neurons in a stochastic model.}
\textbf{a,} An example network represented as an adjacency matrix $A$.
\textbf{b,} A Markov chain of network states can accurately predict the fraction of active cascades at time $t$. In $10^4$ trials of stimulating neuron 8 in the network in panel b, the root-mean-square error between the state-space prediction and the stochastic model is $1.2 \times 10^{-4}$.
\textbf{c,} Examples of simulations of cascades generated by stimulating neuron 8 in the network in panel b.
\textbf{d,} The activity of each node and time step, averaged over $10^4$ cascades of stimulating neuron 8 in the network in panel a.
\textbf{e,} Linear dynamics estimates the average spike counts of stochastic simulations in panel d.
\textbf{f,} The difference between linear dynamics and simulation average converges to a steady-state around zero.
\textbf{g,} The differences between average simulated spiking and estimated linear dynamics for weighted random networks of size 50, 100, 150, 200, 250, and 300 nodes, all with fractional connectivity of 0.2. The error bars indicate standard deviations, and the means are $2.1 \times 10^{-4}$, $-6.8 \times 10^{-5}$, $-9.7 \times 10^{-6}$, $-8.0 \times 10^{-5}$, $5.8 \times 10^{-5}$, and $-2.1 \times 10^{-5}$, respectively.
}
\label{fig1}
\end{figure}

We begin with the stipulation of a network as well as a dynamical process that occurs atop the network. We formalize the notion of a network as a directed graph $\mathcal G = (\mathcal V, \mathcal E)$ in which neurons are represented as nodes $\mathcal V = \{1, \dotsm, n\}$ and neuron-to-neuron connections are represented as edges $\mathcal E \subseteq \mathcal V \times \mathcal V$. The weighted and directed adjacency matrix $A = [a_{ij}]$ thus encodes the edge weights from neuron $j$ to neuron $i$ (Figure \ref{fig1}a).

\subsection*{Stochastic McCulloch-Pitts neuron}
To model neuronal cascades, we next stipulate a stochastic version of the McCulloch-Pitts neuron \cite{mcculloch1943}. In the McCulloch-Pitts model, a neuron receives inputs scaled by the weights of the edges and sums the scaled inputs, $\bm{a}_i \cdot \bm{y}$, to produce an output via an activation function. Here, the activation function is a random Bernoulli process, where probability $p$ is the sum of the scaled inputs. The sum of the scaled inputs $\bm{a}_i \cdot \bm{y}$ is bound by 0 and 1 such that $p=\min(1,\max(0,\bm{a}_i \cdot \bm{y}))$. The network starts at some non-random initial state $\bm{y}(0)$, which can also be interpreted as a stimulus received at $t=0$. The state of an $n$-neuron network is a binary vector $\bm{y}(t) \in \{0,1\}^n$ such that each element indicates whether a neuron fired at time $t$ and evolves as

\begin{align}
\label{eq:bernoulli}
y_i(t) \sim B(\bm{a}_i \cdot \bm{y}(t-1)) \text{,}
\end{align}
where $B(r)$ is a Bernoulli process with probability $r$, and $\bm{a_i}$ is the $i^{th}$ row vector of $A$.

\subsection*{Markov chain formulation}
When $\sum_i a_{ij} \leq 1$, the model that we consider can be represented as a Markov chain with states $\bm{s}^i \in \{0,1\}^n$ representing all possible patterns of spikes in the network, and with state $\bm{s}^1 = \bm{0}$ representing the zero state. The column vector $\bm{p}(t) = [p_1(t); \dotsm; p_{2^n}(t)]=[P(\bm{y}(t)=\bm{s}^1); \dotsm; P(\bm{y}(t)=\bm{s}^{2^n})]$ represents the probability that the network exists in any state $\bm{s}^i$ at time $t$. The transition matrix $T$ governs

\begin{align}
\label{eq:state}
\bm{p}(t) &= T \bm{p}(t-1) = T^t\bm{p}(0) \text{,}
\end{align}
where each entry $T = [T_{lk}] = P(\left[\bm{y}(t) = \bm{s}^l \right] | \left[\bm{y}(t-1) = \bm{s}^k\right])$ represents the transition probability from state $k$ to state $l$. See Methods for details regarding the computation of the matrix $T$.

The Markov representation above makes explicit the relationship between the network $A$ and the stimulus propagation and discrimination. The process stated in Equation \ref{eq:bernoulli} determines a unique map from adjacency matrix $A$ to transition matrix $T$. Given an initial distribution of states, i.e., the stimulus patterns, $\bm{p}(0) = [p_0(0); p_1(0); \dotsm]$, the fraction of cascades that terminate by time $t$ is simply given by the first entry of $\bm{p}(t)$. Conversely, the probability that a cascade is alive at time $t$ is given by $P(\text{alive},t) = 1-p_0(t)$. Similarly, the discrimination between network states propagated from stimulus $\bm{y}(0) = \bm{s}^i$ and from $\bm{y}(0) = \bm{s}^j$ from some measurement $\bm{y}(t)$ depends upon the similarity between probability vectors $\bm{p}_i(t)=T^t\bm{s}^i$ and $\bm{p}_j(t)=T^t\bm{s}^j$. For quickly decaying systems, $\bm{p}_i(t)$ and $\bm{p}_j(t)$ will both have a high probability of being in the zero state $\bm{s}^1$, inherently reducing discriminability. Hence, the architecture of the network $A$ constrains the amount of persisting activity that permits discrimination of the initial spiking distribution $\bm{p}(0)$.

To numerically assess the constraint on cascade duration, we compared simulations of the network in Figure \ref{fig1}a to the prediction $P(\text{alive},t)$ given by the Markov representation and observed little difference between the stochastic and predicted dynamics. For each of $10^6$ trials, we stimulated single neurons at $t=1$, and at each time step (from a maximum of 100), we calculated the fraction of cascades alive and $P(\text{alive},t)$. We found that the root-mean-square error (RMSE) between the Markov chain prediction and the stochastic model was $1.2 \times 10^{-4}$ (Figure \ref{fig1}b). To determine the generalizability of our observations, we extended this analysis to an ensemble of 120 networks, separated into 30 instantiations of four different graph topologies chosen for their relevance to neuronal architectures: a weighted random graph, a ring lattice graph, a modular graph with 4 communities, and a Watts-Strogatz graph (see Methods). For the four graph topologies, we observed that the average RMSEs were less than $8.5 \times 10^{-3}$. Taken together, these results indicate a tight link between network structure $A$ and cascade duration derived from the network dynamics $T$.

\subsection*{Estimation as a linear dynamical system}
Because computing a Markov chain is intractable for large network sizes, we instead estimate the process stated in Equation \ref{eq:bernoulli} with a linear dynamical system with the same parameters $A$. Specifically, the average activity generated by the stochastic model can be written as $\bm{x}(t) = \mathbb{E}[\bm{y}(t)]$, and given equal initial states $\bm{x}(0)=\bm{y}(0)$ and $\forall i \in \mathcal V: \sum_j a_{ij} \leq 1$, it is straightforward to show that this average network state obeys

\begin{align}
\label{eq:mean}
\bm{x}(t) = A\bm{x}(t-1)
\end{align}
(see Methods for a formal proof). Equation \ref{eq:mean} offers a natural intuition: the average behavior of the stochastic model follows linear dynamics and evolves exponentially as a function of time. Such a relationship allows the application of rich mathematical principles of linear dynamical systems to describe average stochastic dynamics of the model.

To illustrate the linear relation, we perform numerical simulations of the model, and we compare simulated cascades to the average network state estimated by a linear dynamical system. In both cases, we instantiate the dynamics on a weighted random network comprised of 10 nodes (Figure \ref{fig1}a) \cite{garlaschelli2009}. The important network parameters for all simulations are listed in the Supplemental Information. We simulate the stochastic model dynamics 1,000 times over 15 time steps starting with the same initial condition $\bm{y}(0)$ (Figure \ref{fig1}c). Note that we can also consider this initial condition to be the stimulus. We average the activity at each node and time step $y_j(t)$ across simulations to generate a numerical estimate of the time-evolution of the average network state (Figure \ref{fig1}d). Then, using the linear dynamical system starting with the same initial condition $\bm{x}(0)=\bm{y}(0)$, we calculated the number of spikes per neuron per time step as an estimate of the average network state (Figure \ref{fig1}e).  We find that the difference between the states of the linear system and of the stochastic cascading model approaches 0 as a function of trials $k$ (Figure \ref{fig1}f). This convergence is consistent across a range of network sizes for fixed density (Figure \ref{fig1}g). These results illustrate the accuracy of the linear estimation of the dynamics of the stochastic model.

\section*{Results}

\subsection*{Network structure constrains cascade duration}
\begin{figure}[!ht]
\centering
\includegraphics[width=\linewidth]{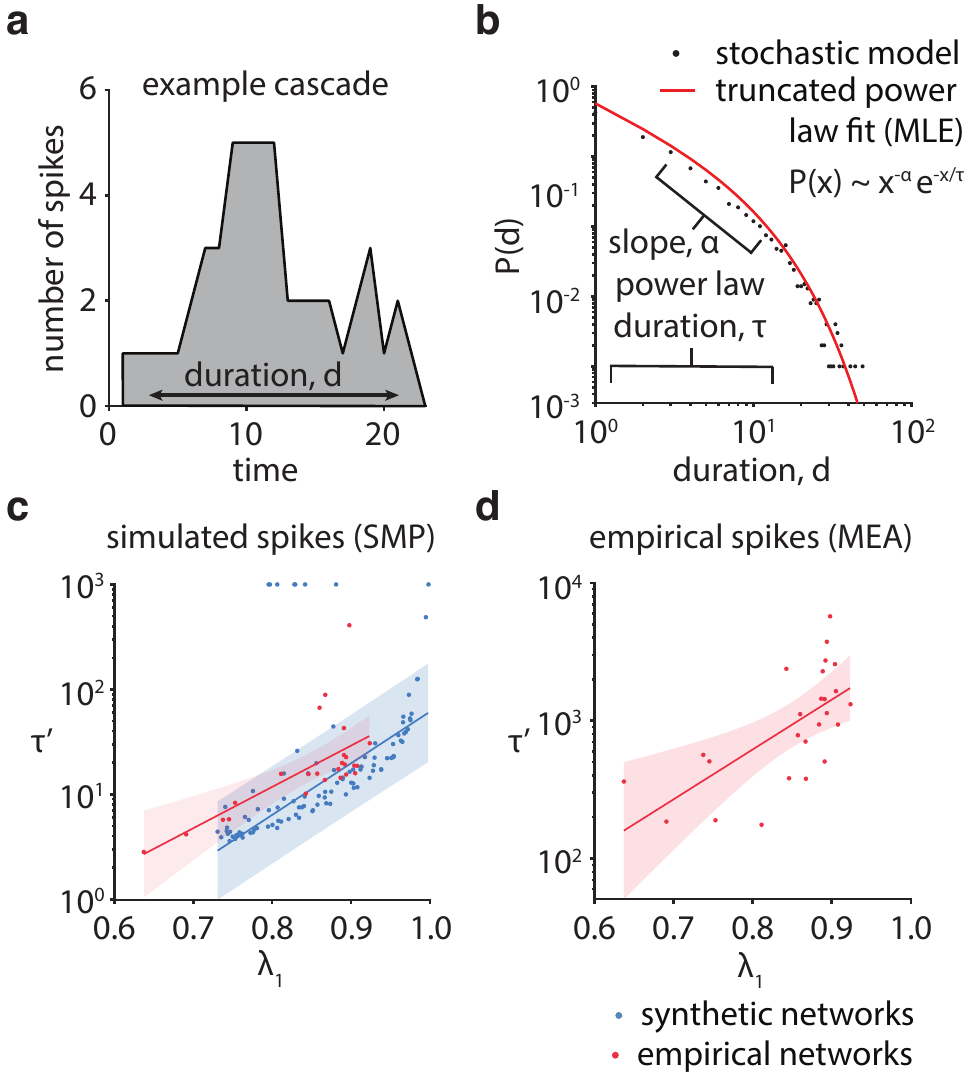}
\caption{
\textbf{Network topology constrains cascade duration.}
\textbf{a,} Cascade duration is defined as the number of time steps $t$ between the point at which the first spike occurs after a time step of quiescence, and the point at which the last spike occurs, followed by a time step of quiescence.
\textbf{b,} The distribution of cascade duration can be described by a truncated power law, where parameter $\alpha$ indicates the log-log slope of the initial distribution and $\tau$ indicates the duration of the power law on the distribution.
\textbf{c,} In simulations of the stochastic McCulloch-Pitts (SMP) model, the dominant eigenvalue $\lambda_1$ of synthetic (blue) and empirical (red) networks monotonically scales $\tau'$ with Spearman's $\rho$ of 0.93 and 0.68 ($p < 0.001$), respectively. Simulations are run for $10^3$ time steps.
\textbf{d,} In 25 multielectrode (MEA) recordings, the dominant eigenvalue $\lambda_1$ of empirical networks monotonically scales $\tau'$ with Spearman's $\rho$ of 0.69 ($p < 0.001$).
}
\label{fig2}
\end{figure}

As the first step towards uncovering relations between architecture and cascading dynamics, we provide a mathematical relationship between network topology and cascade duration using intuitions grounded in the theory of linear dynamical systems (Figure \ref{fig2}a). To more generally describe cascade behavior than in the Markov representation, we can decompose the weight matrix $A$ into eigenvalues and eigenvectors to identify the elementary modes of activity propagation. Using the dominant eigenvalue $\lambda_1$ to identify the constraint on the dominant propagation of activity, we can estimate nonlinear, stochastic behavior with a linear system. The dominant eigenvalue $\lambda_1$ is defined as

\begin{align}
\label{eq:eig}
\lambda_i \in \text{eig}(A): Av_i=\lambda_i v_i \text{,}
\end{align}
with the maximum absolute value. The dominant eigenvalue $\lambda_1$ scales the dominant eigenvector $v_1$, which constrains the most persistent mode, or vector, of activity propagation in the network \cite{seung1996, larremore2011a, larremore2011b}.

To numerically demonstrate the utility of the metric $\lambda_1$ in explaining cascade duration, we simulated cascades on 104 networks with $2^8$ nodes for $10^3$ time steps (see Methods and Supplemental Information for network parameters). Using maximum likelihood estimation (MLE) \cite{clauset2009, alstott2014}, we fit a truncated power law $p(x) \sim x^{-\alpha}e^{-x/\tau}$ to distributions of cascade duration and computed $\tau'$ as $\tau$ bounded by the maximum duration, $\min(d_{\max},\tau)$ (Figure \ref{fig2}b). Intuitively, the metric $\tau'$ captures the temporal scale in which activity can propagate in a network. We found that $\tau'$ is monotonically correlated with $\lambda_1$, with a Spearman's correlation coefficient $\rho$ of 0.93 ($p \approx 0$), and that $\alpha$ has a mean of $2.0 \pm 0.14$ (standard error; Figure \ref{fig2}c) \cite{bak1987}. Notably, these relations can inform how one would tune the network $A$ to produce heavy-tailed distributions of cascade duration.

Finally, we tested our predictions in empirical data and find similar correlations between network structure and dynamics (Figure \ref{fig2}c,d). In each of 25 multielectrode array (MEA) recordings of spiking neurons in the mouse somatosensory cortex \cite{ito2016}, we binned the spikes into 5ms bins and used MLE to fit a truncated power law and compute $\tau'$. To derive $\lambda_1$, we calculated an effective connectivity matrix from each recording using first-order vector autoregression (VAR) \cite{neumaier2001, schneider2001}. We found that $\tau'$ is monotonically correlated with $\lambda_1$, as reflected in a Spearman's $\rho$ of 0.69 ($p=1.8 \times 10^{-4}$; Figure \ref{fig2}d). Moreover, we can simulate stochastic cascades on the empirically derived networks and find a significant positive correlation between $\tau'$ and $\lambda_1$, with a Spearman's $\rho$ of 0.68 ($p=2.6 \times 10^{-4}$; Figure \ref{fig2}c). With a mean $\alpha$ of $2.3 \pm 0.1$, these recordings range in their proximity to criticality (see Supplemental Information for their exponent relations), yet their dynamics are all well-described by their network structures. All together, these results demonstrate the dependence of the temporal scale of activity propagation on the network structure of neural systems.

\subsection*{Local network structures: cycles and connection strength}
Having demonstrated in the previous section that cascade duration can be predicted from the network structure, we next turn to a deeper examination of which specific features of a network's topology and geometry can support a heavy-tailed distribution of cascade duration. Note that we use the phrase \textit{network topology} to indicate the arrangement of binary edges and we use the phrase \textit{network geometry} to indicate the distribution of edge weights \cite{bassett2013}. The two candidate features that we consider are (i) the presence of cycles and (ii) the strength of connections in cycles. We will study these features through a rewiring process on an initial set of edges.

\begin{figure*}[!ht]
\centering
\includegraphics[width=\textwidth]{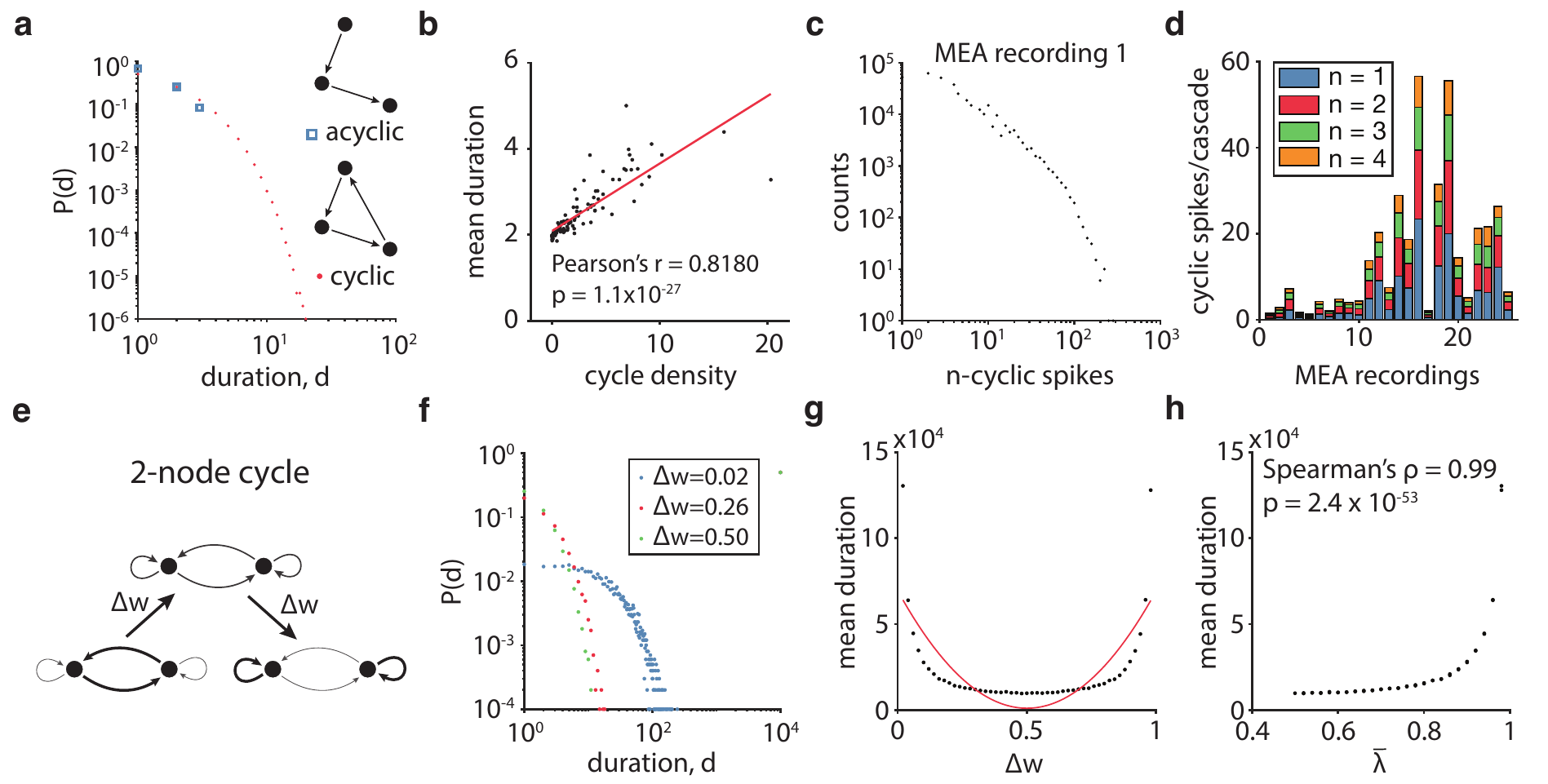}
\caption{
\textbf{Cycles and strong connections facilitate long cascades.}
\textbf{a,} Distribution of cascade duration in acyclic and cyclic networks (all edges have weight 0.5). In $10^6$ trials, the maximum cascade durations for the acyclic and cyclic networks are 3 and 11 time steps, respectively.
\textbf{b,} Networks with higher cycle density have longer cascades. We randomly rewired a directed acyclic graph to produce networks of varying cycle density. Cycle density is the number of simple cycles divided by the number of edges.
\textbf{c,} Distribution of $n$-cyclic spikes in MEA recording 1 with 5ms bins. An $n$-cyclic spike occurs when node $i$ fires and then fires again after $n$ time bins.
\textbf{d,} Neurons in mouse somatosensory cortex fire in cycles with small refractory periods. Plot shows distribution of the average number of $n$-cycles observed per cascade ($9.3 \times 10^4 \pm 6.8 \times 10^3$ cascades for 25 recordings; standard error).
\textbf{e,} A schematic of a 2-node network. We redistributed the weights from the 2-node cycle to self-loops by $\Delta w$.
\textbf{f,} Distributions of cascade duration for $\Delta w = 0.02$, $0.26$, and $0.50$ in the 2-node cycle.
\textbf{g,} Cycles with strong connections, at either $\Delta w \rightarrow 0$ or $\Delta w \rightarrow 1$, extend the mean duration of cascades that do not reach fixed point $\bm{1}$ (quadratic fit: $y=(2.7\times10^5)x^2-(2.7\times10^5)x+(6.9\times10^4)$).
\textbf{h,} Mean eigenvalue $\bar{\lambda}$ can characterize a network geometry's capacity for long duration of cascades that do not reach fixed point $\bm{1}$.
}
\label{fig3}
\end{figure*}

\textit{The prevalence of cycles.} We begin by noting that cycles support temporally extended cascades. Given a single initial stimulus or spontaneous spike, a cascade can have a duration greater than the number of nodes in the graph if and only if there exists at least one cycle in the network. We demonstrate this simple intuition with an acyclic 3-node network and a cyclic 3-node network, where each edge in both networks has a weight of 0.5 (Figure \ref{fig3}a). In simulations of $10^4$ cascades, we found that the acyclic network produces a maximum cascade duration of 3 time steps, as expected. In contrast, using the same number of simulations on the cyclic network, we found the much greater maximum cascade duration of 13 time steps.

Next, we show that the cascade duration scales monotonically with the prevalence of cycles in a network as measured by cycle density, which we define as the number of simple cycles divided by the number of connected edges (Figure \ref{fig3}b). To study the effect of cycle density, we begin with a 10-node, directed acyclic graph and randomly rewire each edge with probability $p$ to a different target node. The directed acyclic graph has the maximum number of edges; that is, the weight matrix is an upper triangular matrix without the diagonal entries. By sweeping over rewiring probabilities $p = \{0.0, 0.1, 0.2, ..., 1.0\}$, we generated networks with different numbers of simple cycles, but the same number of edges and same edge weights of $\frac{1}{n}$. For each $p$, we simulated $10^4$ cascades with a maximum duration of $10^4$, and we measured the slope of the linear tail of the distribution on a log-log plot. In these simulations, we found that as a network is rewired to contain more cycles, the average cascade duration increases (Pearson's correlation coefficient $r=0.8180$, $p = 1.0835\times 10^{-27}$; Figure \ref{fig3}b). These examples illustrate the more general rule that networks containing cycles can support longer cascades and can extend the tail of the distribution of cascade duration.

Importantly, we empirically validate that activity can propagate through cycles. A key potential constraint for cyclical activity propagation is a large refractory period, which can impede such activity even if cycles are structurally present \cite{vankessenich2016}. Hence, using the same 25 MEA recordings of spiking cortical neurons as employed previously, we measured the extent of cyclical activity by quantifying the occurrence of $n$-cyclic spikes, a phenomenon which occurs in a cascade when a neuron spikes again after $n$ time bins of its previous spike. We found that on average, 1-, 2-, 3-, and 4-cyclic spikes occur $14.5 \pm 3.1$ times per cascade (with an average of $(9.3 \times 10^4) \pm (6.8 \times 10^3)$ cascades for 25 recordings, standard errors; Figure \ref{fig3}c,d). With a 5ms bin width, these cyclical activity patterns are within biophysical limits \cite{connors1990}. Collectively, these result suggest that cyclical activity propagation is not impeded by refractory periods and indeed occurs frequently in living neuronal systems.

\textit{The strength of connections in cycles.} We now turn to a consideration of the distribution of edge weights. To maximize the specificity of our inferences and to generally build our intuition, we constrained ourselves initially to simple networks that only contain a small cycle (a 2-node cycle) or that also contain one relatively larger cycle (a 4-node cycle; see Supplemental Information). We probed the role of weight distributions in the dynamics of the network by placing the strongest weights on edges on one cycle and by placing the weakest weights on edges not on that cycle. Specifically, in both the 2-node and 4-node cycle networks for each simulation, we took the strong weights initially placed on the cycle and redistributed some of their weight by $\Delta w$ to randomly chosen edges that are not part of the original cycle (Figure \ref{fig3}e). Upon these new networks, we simulated the stochastic model. We found that as the weight on the original cycle is continuously redistributed away from the initial cycle and throughout the network, we observe fewer and fewer cascades of long duration (Figure \ref{fig3}f,g).

Across empirical studies \cite{beggs2003, petermann2009, hahn2010, friedman2012, poil2012, lombardi2014, bellay2015, shew2015, poncealvarez2018}, the distributions of avalanche duration have been described by power law functions, where the exponent is known as the lifetime. Typical values vary from -1.0 to -2.6. We seek to show how cycle density and edge weights in cycles together explain the topological and geometric differences in the networks underlying the various distributions of cascade duration. As we redistributed edge weight more uniformly in the networks, we found that mean duration of terminated cascades increases (Figure \ref{fig3}f,g). Furthermore, as we redistributed away from the uniform geometry, continuously increasing the range of edge weights, we again observed more and more cascades of long duration. These observations underscore the tight coupling between the range of edge weights, and the heavy-tailed nature of the distribution of cascade duration.

Lastly, we seek to determine whether the distribution of edge weights along cycles contributing to cascade duration is captured by eigenvalue analysis. Towards this goal, we employed the same perturbative numerical experiments on the networks. Specifically, we found that as the weights of the original cycle are redistributed, the mean duration of cascades tracks monotonically with the average of eigenvalues of the network (Spearman's rank correlation coefficients $\rho=0.99$, $p=2.4\times 10^{-53}$ and $r=0.46$, $p=9.0 \times 10^{-4}$, respectively; Figure \ref{fig3}h). Because the dominant eigenvalues of the networks in these simulations are all equal to 1, the average of eigenvalues provide a more descriptive estimation of activity propagation. Thus, this result suggests that edge weight constrains cascade duration by determining the strength of activity propagation.

\subsection*{Node-specific dynamics}
Even within a single network architecture, the range of cascade dynamics can vary depending on the nodes that are stimulated, either spontaneously as the initial state of a cascade or exogenously through input. Thus, we now consider the role of the stimulus pattern on cascade dynamics. We extend our eigenvalue analysis to estimate the role of a stimulus pattern on stochastic cascade dynamics by calculating the magnitude of the eigenprojection of the stimulus pattern. Because the average dynamics are explained by linear systems theory, we then use network control theory to more accurately predict how stimulation of individual nodes alters the dynamics of cascades.

\textit{The eigenprojection of the stimulus pattern.} First, we tested whether the magnitude of the eigenprojection of the stimulus pattern could predict cascade dynamics. Given a stimulus $\bm{y}(0)$, the eigendecomposition of the weight matrix $A$ into $A=PDP^{-1}$ yields $\bm{c} = P^{-1}\bm{y}(0)$ as the coefficients of the eigenmode excitation of $\bm{y}(0)$. The components of $\bm{c}$ determine how much the stimulus $\bm{y}(0)$ projects onto the eigenvectors of $A$ and describes the modes of average activity propagation through the network $A$. Then, as a predictor for mean duration, we can compute the 1-norm, or the sum of absolute values, of the eigenprojection of the stimulus pattern scaled by the corresponding eigenvalues $\bm{\lambda}$,

\begin{align}
\label{eq:magproj}
|\bm{c} \cdot \bm{\lambda}|_1 \text{.}
\end{align}

\begin{figure}[!ht]
\centering
\includegraphics[width=\linewidth]{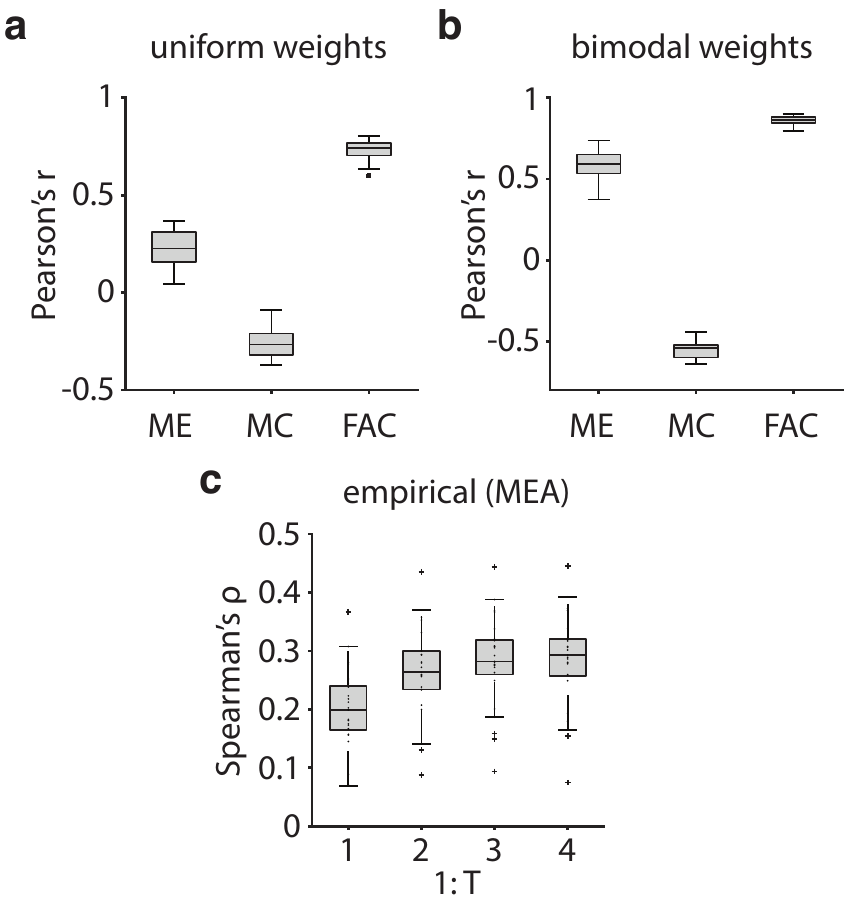}
\caption{
\textbf{Network controllability is tightly linked with cascade duration.}
\textbf{a,} Pearson's correlation coefficients between mean duration of cascades from the stimulation of individual nodes and controllability measures of the respective nodes. The controllability measures include the magnitude of the eigenprojection (ME), modal controllability (MC), and finite average controllability (FAC). The networks here are 30 random instantiations of weighted random graphs, each with 100 nodes and a fractional connectivity of around 0.2.
\textbf{b,} The same plot as in panel a except with a bimodal distribution of weights---with 10\% of connections normally distributed with a mean of 0.9 and 90\% of connections with a mean of 0.1, all with a standard deviation of 0.1, before weight normalization.
\textbf{c,} Controllability measurements in spiking neurons in mouse somatosensory cortex predict cascade duration. Spearman's correlation between the duration of each cascade and mean finite average controllability of neurons active in its first $T$ time bins (5ms bins) for 25 MEA recordings. See Supplemental Information for individual plots. The box-plot elements, center, bottom and top edges, whiskers, ``+" symbols, indicate respectively, the median, 25th and 75th percentiles, extremes, and outliers.
}
\label{fig4}
\end{figure}

We numerically test the eigenprojection metric by simulating cascades on a 100-node, weighted random network. The mean duration of cascades generated from the stimulation of a single node was significantly positively correlated with the magnitude of the eigenprojection (Pearson's correlation coefficient $r=0.34$, $p = 4.5 \times 10^{-4}$). To determine the generalizability of these findings, we expanded our simulation set to include 30 random instantiations of networks with the same parameters. In this broader dataset, we found that the Pearson's correlation coefficient was highly variable (median $r=0.23$; Figure \ref{fig4}a). Thus, we can weakly estimate the role of a stimulus pattern on cascade dynamics with eigenvalue analysis.

\textit{Network control theory.} To more accurately predict the role of a stimulus pattern on cascade dynamics, we adopt the recently developed metrics of average and modal controllability from network control theory \cite{pasqualetti2014controllability}. We hypothesized that these metrics, previously applied to large-scale brain networks \cite{gu2015, tang2017developmental}, predicts cascade duration since network control necessitates activity. In the same set of simulations reported above, we compared the mean cascade duration to the finite average controllability of each node, defined as

\begin{align}
\text{Trace}(W_K) \text{,}
\end{align}
where $W_K=\sum_{\tau=0}^F A^\tau B_K (A^\tau B_K)^\top$ is the finite controllability Gramian (see Methods). We observed that the mean cascade duration and finite average controllability were significantly positively correlated (Pearson's correlation coefficient $r=0.79$, $p = 2.7 \times 10^{-22}$). In contrast, modal controllability was not strongly correlated with mean cascade duration (Pearson's correlation coefficient $r=-0.12$, $p = 0.24$). To determine the generalizability of these findings, we expanded our simulation set to include 30 random instantiations of networks with the same parameters. In this broader dataset, we observed consistent effects (median Pearson's correlation coefficient $r=0.74$ and $r=-0.27$ for finite average controllability and modal controllability, respectively; Figure \ref{fig4}a). In comparing the predictions from linear control theory with the predictions from eigendecomposition, we note that finite average controllability is consistently more strongly correlated with the mean cascade duration than the magnitude of the eigenprojection.

Interestingly, networks with the same topological parameters as above, but with a bimodal distribution of weights show even stronger correlations between network control statistics and cascade dynamics (Figure \ref{fig4}b). Such a weight distribution reduces variance in the stochastic process, which intuitively can serve to strengthen the correlation. We observed that the mean cascade duration and finite average controllability were significantly positively correlated (Pearson's correlation coefficient $r=0.87$, $p = 3.2 \times 10^{-32}$). Modal controllability became strongly negatively correlated with mean cascade duration (Pearson's correlation coefficient $r=-0.50$, $p=9.2 \times 10^{-8}$). Again to determine the generalizability of these findings, we expanded our simulation set to include 30 random instantiations of networks with the same parameters. We observed consistent effects (mean Pearson's correlation coefficients between mean cascade duration and finite average controllability, modal controllability, and magnitude of eigenprojection were $r=0.86$, $r=-0.54$, and $r=0.59$, respectively; Figure \ref{fig4}b). Again we note that finite average controllability is consistently more strongly correlated with the mean cascade duration than the magnitude of the eigenprojection. These simulations suggest that the skewed weight distributions, as identified in the previous section as network motifs that support long cascades, may strengthen the relationship between network control and network dynamics. Collectively, the results illustrate that the stimulus patterns and the network must be tailored for each other to produce the desired neural dynamics. Our observations naturally lead to the question of how stimulation, either endogenous or exogenous, can be used for information processing.

Finally, we tested these predictions in empirical data and find that controllability of the initial states is correlated with cascade duration (Figure \ref{fig4}c). In each recording from the same MEA data used earlier from spiking neurons in the mouse somatosensory cortex, we calculated the mean finite average controllability of all nodes active in the first $\{1...T\}$ time bins of each cascade. Mean finite average controllability is monotonically correlated with the duration of each cascade with a median Spearman's $\rho=0.20$ for $T=1$ and $\rho=0.26$ for $T=2$ ($p<0.001$, Bonferroni-corrected). It is important to remember that the cascades are stochastic and cannot be predicted deterministically. Thus, it is notable to find any correlation between mean finite average controllability and cascade duration in empirical data.

\subsection*{Cascade duration allows network discriminability and stimulus recovery}
If certain network topologies and stimulus patterns can produce long-lasting cascades consistent with avalanche dynamics, what role can lasting cascades contribute to information processing? Intuitively, one cannot recover information about stimuli from cascades that have already terminated. For lasting cascades, network states can be discriminated and can also provide information about stimuli. Such delayed recovery of stimuli can allow the associative learning of stimuli across temporal delays \cite{goldmanrakic1995, durstewitz2000, eriksson2015}. The intuition that lasting cascades allow network discriminability can be formalized mathematically via Equation \ref{eq:mean}. Then, with simulations, we test the intuition that cascade dynamics support stimulus recoverability.

\textit{Network discriminability.} To analytically show the relationship between cascade duration and discriminability, we first define network discriminability as the Euclidean distance between two states $d(\bm{y_1}(t),\bm{y_2}(t))$ in $n$-dimensional space. Recall that $\mathbb{E}[\bm{y}(t)] = \bm{x}(t)$ for stimulus $\bm{x}(0)$ from Equation \ref{eq:mean}. Then, given two stimuli, $\bm{x_1}(0)$ and $\bm{x_2}(0)$, we can calculate the expected network discriminability as the distance between the expected network states $d(\bm{x_1}(t),\bm{x_2}(t))$ at time $t$. Given that the dominant eigenvalue $\lambda_1 < 1$, then $\bm{x}(t)$ approaches the zero vector $\bm{0}$ as $t$ approaches $\infty$. As described in previous sections, the decay in activity is constrained by the dominant eigenvalue of the network and by the finite average controllability of the individual node being stimulated. Thus, the rate at which both $\bm{x}_1(t)$ and $\bm{x}_2(t)$ decay to $\bm{0}$ determines the rate at which $d(\bm{x_1}(t),\bm{x_2}(t))$ approaches $d(\bm{0},\bm{0})$ where discriminability between two network states is zero.

\begin{figure*}[!ht]
\centering
\includegraphics[width=\textwidth]{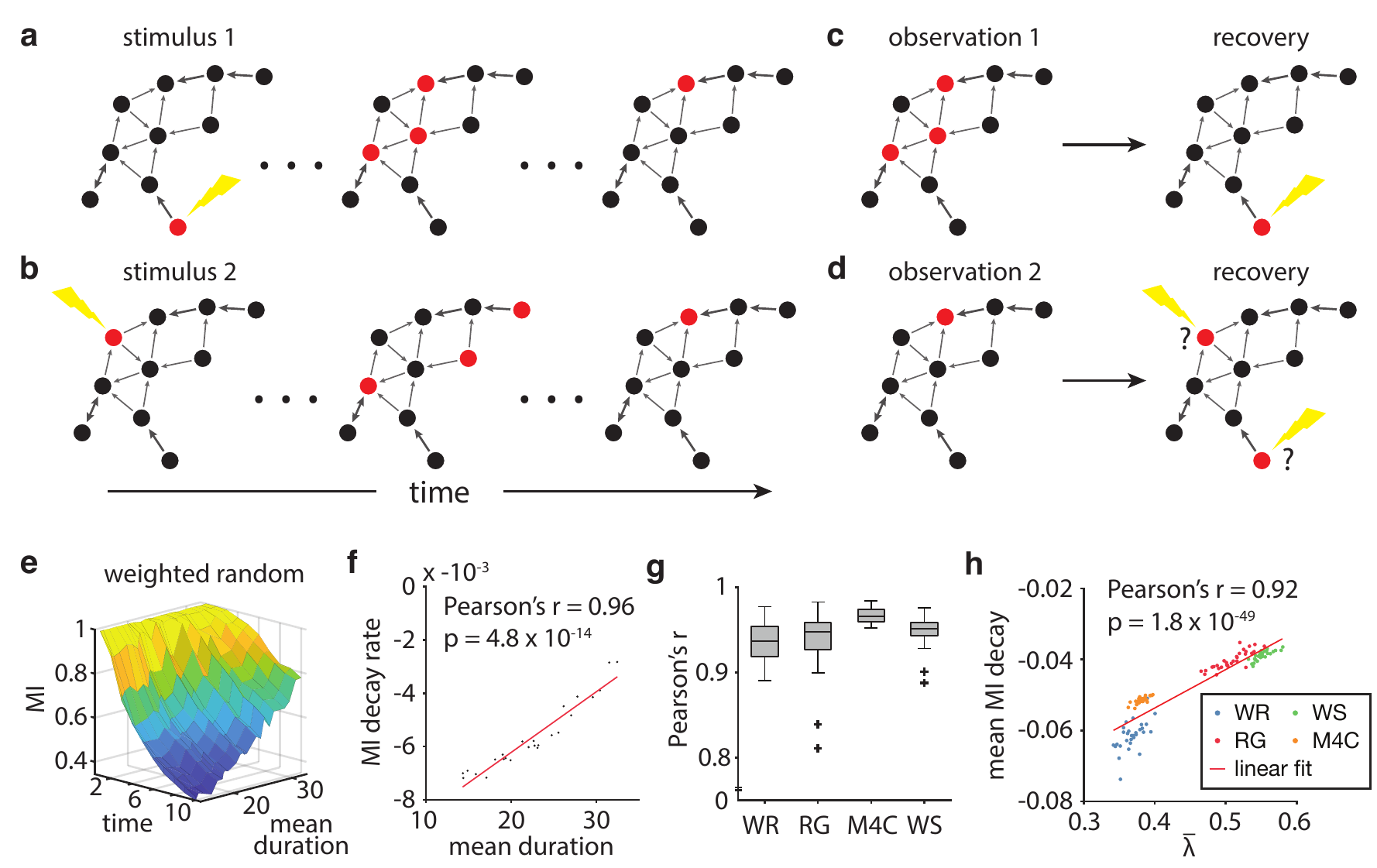}
\caption{
\textbf{A stimulus can be well-recovered when it generates long-lasting cascades.}
\textbf{a-b,} A schematic showing two cascades triggered by different stimuli.
\textbf{c,} Recovery of the stimulus using an observation of a network state during a cascade.
\textbf{d,} Failed recovery of the stimulus.
\textbf{e,} Decay in mutual information (MI) over time. When activity from a stimulus pattern lasts longer, mutual information also persists for longer for a weighted random graph.
\textbf{f,} The linear decay rate of mutual information over the first 10 time steps plotted against the mean cascade duration in the example weighted random graph from panel e.
\textbf{g,} The Pearson correlation coefficients between the linear slope of decay in mutual information over time and the mean cascade duration for four graph types: a weighted random graph (WR), a random geometric graph (RG), a modular graph with 4 communities (M4C), and a Watts-Strogatz graph (WS). The boxplot shows data from 30 instantiations of each graph type, each network containing 100 nodes and characterized by a fractional connectivity of around 0.05. The whiskers extend to the extreme data points not considered outliers, and the outliers are plotted individually using the ``+" symbol.
\textbf{h,} The mean decay rate in mutual information for a network is correlated with the sum of eigenvalues of the network (Pearson's correlation coefficient $r=0.92$, $p=1.8 \times 10^{-49}$). For all networks, we used a fractional connectivity of 0.05 to show a wide range of decay rates in mutual information (see Supplemental Information for simulations with other fractional connectivities).
}
\label{fig5}
\end{figure*}

\textit{Stimulus recovery.} To numerically show the relationship between cascade duration and stimulus recoverability, we first define stimulus recoverability as the mutual information $I(S;Y_t)$ between stimulus patterns $s \in S$ and network states $y \in Y_t$ at time $t$ (see Methods for details and Figure \ref{fig5}a-d for an intuitive schematic). Similar to discriminability, mutual information between the stimuli and network states decreases with shorter cascade duration because the Shannon entropy of the network states decreases. To probe this relation formally, we simulated cascades with 100-node networks from 4 different graph topologies with 30 instantiations of each graph type. Consistent with our intuition, we observe that mutual information is maintained longer when cascades last longer on average (Figure \ref{fig5}e). We then quantified the decay in mutual information by first performing linear regression on the mutual information as a function of time for the first 10 time steps. By calculating the Pearson's correlation coefficient between the slope of linear regression and the mean cascade duration, we found that for all four graph topologies, mutual information decays faster when the propagation of activity also decays faster (Figure \ref{fig5}f). Collectively, these results demonstrate that stimulus recoverability is maintained longer when the cascades generated by stimulus patterns last longer.

To link information retention back to network structure, we assessed the relation between stimulus recoverability and the sum of eigenvalues of each network. Using the same 100-node networks from 4 different graph topologies with 30 instantiations of each graph type, we found a significant positive correlation between the average decay rate in mutual information and the sum of eigenvalues, implying that network structure supports the retention of information within the network (Pearson correlation coefficient $r=0.92$, $p=1.8 \times 10^{-49}$; Figure \ref{fig5}g). Moreover, while all networks had similar parameters, each graph type generated distinct ranges of decay rates and sums of eigenvalues, suggesting that certain graph types may be better suited for information retention than others (Figure \ref{fig5}h). In particular, we observe lower decay rates in mutual information and lower sum of eigenvalues in the weighted random and modular graphs, than in the random geometric and Watts-Strogatz graphs. Collectively, these findings demonstrate the interplay among network architecture, network dynamics, and information processing.

\section*{Discussion}

Neural systems display strikingly rich dynamics that harbor the marks of a complex underlying network architecture among units, from the small scale of individual neurons to the large scale of columns and areas \cite{wang2013, nigam2016}. Cascades are a quintessential example of such dynamics, and, when they display features of self-organizing criticality, are thought to allow for a diverse range of computations \cite{beggs2003, haldeman2005, kinouchi2006, shew2009}. Yet, precisely how a neuronal network's structure supports stochastic dynamics and the computations that can arise therefrom remains unclear. Here, we seek to provide clarity using both precise analysis of mathematical formulations and statistically rigorous assessments of numerical experiments. We consider a generalized stochastic spiking model and demonstrate that the time-averaged activity of this model can be treated as a linear dynamical system. From this observation, we derive intuitions for how network structure constrains cascade duration. In subsequent numerical experiments and empirical validation, we use eigendecomposition and statistical approaches from network control theory appropriate for linear dynamical systems to describe how network structure and the stimulus pattern together determine the manner in which a stimulus propagates through the network. We identify strongly connected cycles as prevalent network motifs that promote long cascade duration in neuronal networks. Finally, we use mutual information to demonstrate that long-lasting cascades can serve as a mechanism to allow for temporally delayed recovery of desired patterns of stimulation. Broadly, our work blends dynamical systems theory, network control theory, information theory, and computational neuroscience to address the wide gap in the field's current understanding of the relations between architecture, dynamics, and computation.

\subsection*{Linear form of stochastic network dynamics}
Because of the inherently stochastic nature of neuronal cascades, many previous studies have simply inferred properties about the underlying network through statistical methods \cite{beggs2003, lombard2012}. An important innovation in this study was the demonstration that the time-averaged activity of the stochastic system has an equivalent form as a linear dynamical system. In real neuronal systems, dynamics are non-linear, which most likely accounts for the difference in range of $\tau'$ in Figures \ref{fig2}c and \ref{fig2}d. Such linear estimation of the dynamics makes available powerful computational tools in matrix and linear systems theory, and allowed us to capitalize on recent advances in network control \cite{liu2011, pasqualetti2014controllability}. Network control theory is a formal approach to modeling, predicting, and tuning the response of a networked system to exogenous input, and has been recently applied to neural systems at both the cellular \cite{yan2017, wiles2017autaptic, towlson2018celegans} and regional \cite{gu2015, tang2017developmental, cornblath2018sex, jeganathan2018fronto} scales (for a recent review, see \cite{tang2018control}). In these previous efforts, linear dynamics have been assumed, whereas here such dynamics have been proven, to be relevant for the neural system under study. Extensions of linear systems analysis, such as observability \cite{chen1998} and optimal control \cite{taylor2015optimal, betzel2016optimally, gu2017optimal}, follow immediately from this work and could provide added insights into other dynamical and computational properties of neural networks. Finally, it would be of interest to directly probe the effects of stimulation patterns defined by network controllability statistics on information transmission \textit{in vitro} or behaviors \textit{in vivo}, following work in a similar vein in large-scale human neuroimaging \cite{medaglia2018network, muldoon2016stimulation, stiso2018white, khambhati2018predictive}.

\subsection*{Topological constraints on dynamics and computation}
Proving formally that network topology affects dynamics and computation is important, but can be further complemented by providing intuitions regarding the specific features of a network topology that are most relevant. The identification of functionally relevant features of networked systems has a long history in molecular biology \cite{alon2007network}, with notable efforts identifying structural motifs in transcription regulation networks \cite{shenorr2002network}, protein-protein interaction networks \cite{yeger2004network}, and cellular circuits \cite{hart2012design}, which are thought to arise spontaneously under evolutionary pressures \cite{kashtan2005spontaneous}. Significantly extending prior statistical efforts in large-scale connectomes \cite{sporns2004motifs}, here we demonstrate that specific structural motifs in the form of strongly connected cycles are topological features that support long cascade dynamics. These structural motifs form elementary units or building blocks of the network that can be combined to create connectivity architectures that produce certain dynamical behaviors \cite{shimono2014, nigam2016}. Other theoretical studies have also found strongly and bi-directionally connected neurons as motifs that produce long-lasting memory \cite{brunel2016}, potentially as a mechanism for attractor dynamics \cite{hopfield1982}. Importantly, empirical studies have shown that the network motifs identified here are observed in both cortical microcircuits \cite{wang2006, lefort2009, ko2011, markram1997, song2005, perin2011} and macrocircuits \cite{sizemore2018cliques}. Future work is needed to better understand the rules by which neurons connect to one another, and to determine whether those rules serve to increase the memory capacity of cortical networks. It would also be interesting in the future to determine whether higher-order structural motifs, such as those accessible to tools from algebraic topology \cite{giusti2016twos,sizemore2018importance}, might also play a role in the relationships between topology, dynamics, and computation \cite{sizemore2018cliques, reimann2017cliques}.

\subsection*{Information theory as a performance measure}
To measure information retention, we use mutual information between stimulus patterns and network states. Mutual information, originally developed to study communication channels \cite{shannon1948}, has proven to be a powerful tool for the study of information transmission in avalanching neural networks \cite{beggs2003, shew2011}. While previous studies of neuronal avalanches use power law statistics that suggest criticality as the theoretical link between dynamics and information processing \cite{beggs2003, bertschinger2004, haldeman2005, kinouchi2006, shew2009, shriki2013, shew2015}, we take a more mechanistic approach embedded in dynamical systems theory to study the relationships between network structure, dynamics, and mutual information. In light of recent evidence for the subcriticality of cortical networks and the difficulty of establishing criticality from power laws \cite{priesemann2014, touboul2010, touboul2017}, the direct approach that we take here may prove useful in future studies. Nevertheless, we also acknowledge that our approach has some limitations. Despite its utility in studying information channels, mutual information is unlikely to be the only useful performance measure for a neural system, given the numerous purported computations of cortical networks \cite{shew2009, timme2016}. Indeed, the explanation posited here for the prevalence of strongly connected neurons does not account for the information faculties of the rest of the neural system. Such considerations compel further investigation into how network structure supports other types of information processing accessible to other information theoretic measures.

\subsection*{Methodological considerations}
A few remarks are warranted on the topic of linear dynamics in neural systems. Linear dynamics accurately predicts stochastic, cascade dynamics, and its rich mathematical properties have been used to study neural dynamics in many organisms across a wide range of temporal and spatial scales \cite{liu2011, gu2015, kim2017, yan2017}. At the neuronal level, however, neural dynamics are non-linear \cite{hodgkin1952}. Efforts analytically demonstrating properties about non-linear systems are more limited \cite{motter2015}, and thus, further study is required to more thoroughly demonstrate the relationships shown here in a non-linear system.

\subsection*{Future directions}
In closing, we note that the natural direction in which to take this work will be to consider other types of information processing and to identify network structures and neuronal dynamics of different cell types that produce complex network dynamics which in turn support such computations. Here, we demonstrate that the rich mathematical properties of linear systems can reveal insights into the complex dynamics of non-linear, non-deterministic neural systems. In the future, we can further apply this theory to cascading and other neural systems to ask questions about networks, their dynamics, and their computations. It would be apt to apply this framework to cortical networks from functional, structural, and effective connectivities and to measure memory performance in terms of the network topology and dynamics. It would be interesting to measure differences in memory performance across brain regions, and to test for relationships between topological features and performance. Third and finally, studying well-known network learning rules---such as Hebbian plasticity \cite{hebb1949} and spike-timing dependent plasticity \cite{song2000}---in a dynamical systems and information theoretic framework may shed further light on the functional purpose of these rules.

\section*{Methods}

\subsection*{Synthetic network generation}
We use five different commonly studied graph models from network science in our analyses \cite{wuyan2018}. The first graph model is the \textit{Weighted Random Graph} model (WRG), which is a weighted version of the canonical Erd\"os-R\'enyi model. The weight of an edge is distributed as a geometric distribution with probability of success $p$. Second, we use a \textit{Random Geometric} model (RG) that is embedded in a unit cube, where the edge weights are equal to the inverse of the Euclidean distance between two nodes. We kept only a fraction of the shortest edges in order to achieve a desired edge density $p$. Third, we use a \textit{Modular Graph with 4 Communities} model (MD4). Pairs of nodes within communities have an edge density of 0.8, and nodes across communities are connected to achieve a desired edge density of $p$. The edges of nodes in the same community and across communities are weighted according to a geometric distribution with probability of success $p$ and $1-p$, respectively. Fourth, we use a \textit{Watts-Strogatz} model (WS). The model builds a ring lattice and then uniformly rewires the network, creating a small-world architecture with a random probability of $r=0.1$. Fifth, we use a \textit{Hierarchical Modular Graph} (HM). The model generates a directed network with $m$ hierarchical levels of modules with size $s$, and connection density decays as $1/E^n$. See Supplemental Information for a summary of the graph models used in simulations.

\subsection*{Empirical network generation}
We derive empirical networks by calculating the effective connectivity of spiking neurons in the mouse somatosensory cortex \cite{ito2016}. The data contain 25 recordings, most of which possess hundreds of neurons (min: 98, max: 594, mean: 309, total: 7735). The recordings were performed by multielectrode arrays (MEAs), each with 512 electrodes on a roughly 1mm-by-2mm area. On the spike trains of each recording, we first bin the spike trains into 5ms bins to capture action potential propagation and synaptic transmission in the array area \cite{friedman2012, shimono2014, nigam2016}. We then perform vector autoregression (VAR) to derive an effective connectivity network \cite{neumaier2001, schneider2001}. We set the lower and upper bounds, $p_{min}$ and $p_{max}$ of the model order to 1 and 4, respectively. After selecting an optimal model order $p_{opt}$ using Schwarz's Bayesian Criterion \cite{schwarz1978}, we compute the effective connectivity $A$ as the sum of the coefficient matrices $A_1,...,A_p \in \real^{n \times n}$ of the VAR model over the model orders, $a_{ij}=\sum_p^{p_{opt}} a_{p,ij}$.

\subsection*{Network analysis}
We use three sets of weight distributions: a uniform distribution, a truncated normal distribution, and a bimodal distribution. In some simulations, however, we explicitly set the weights to particular values. In a uniform distribution of weights, we set all weights equal to 1 and normalize each row. In a truncated normal distribution, we set the non-zero weights to the upper half of a truncated normal distribution. A truncated normal distribution of weights has been widely observed both in a theoretical context with synaptic plasticity and in the experimental literature \cite{brunel2004, iyer2013, pehlevan2017}. Lastly, we use a skewed, bimodal distribution with a few connections centered at a normal distribution with a large mean and most other connections centered at a normal distribution with a small mean. Bimodal distributions occur theoretically in the context of additive synaptic plasticity \cite{vanrossum2000}, and positively skewed distributions have been observed experimentally \cite{markram1997a, feldmeyer1999, chen2010}. Our skewed, bimodal distributions combine these two observations by having a few strong connections. All weights are static and do not change with time $t$. See the Supplemental Information for the parameters of the weight distributions for networks used in simulations.

To calculate the cycle density of a graph, we compute the number of simple cycles divided by the number of connected edges. A simple cycle is defined as the set of edges in a closed walk with no repetitions of vertices and edges, other than the starting and ending vertex. The number of simple cycles was calculated using the NetworkX software package (version 2.1) on Python (version 3.7.3).

\subsection*{Simulating the Stochastic McCulloch-Pitts model}
We model cascades as spikes propagating through a recurrent network (see Mathematical Framework). For computational tractability, we set a maximum time step $K$ for the simulations. The simulated spike counts $\bm{y}(t)$ are stored as a $n$-by-$K$ matrix. All simulations and calculations were run on MATLAB (version 2018a) provided by The MathWorks, Inc.

\subsection*{Stimulus pattern generation}
We investigate the propagation of activity through a network initiated by stimulus patterns. The stimulus pattern is set as the initial state $\bm{y}(0)$ or $\bm{x}(0)$ of a network and then propagated forward in time according to either stochastic or linear dynamics, respectively. In our study, we consider two ways to generate stimulus patterns. In the analysis of cascade duration and controllability, we stimulate individual nodes by creating a set of vectors in which the $i^{th}$ element of the $i^{th}$ vector is set at 1 and all other elements are set at 0. In the mutual information analysis, we create a set of column vectors such that their finite average controllability values evenly span the range of controllability values (see later section of this Methods for definition of finite average controllability). In each of the $\frac{n}{m}=25$ vectors, we choose $m=4$ nodes from $n=100$ total nodes to stimulate such that each node that we select is increasing in its finite average controllability value. Because finite average controllability is highly correlated with cascade duration, such input vectors will evenly span the possible duration of cascades.

\subsection*{Predicting cascade dynamics}
We can predict the exact fraction of cascades alive at time $t$ by computing a state transition matrix from any state $k$ to any state $l$. For $k, l \in \{1,...,n\}$, the state transition matrix $T \in \mathbb{R}^{l \times k}$ can be constructed by

\begin{align*}
  &P\left(\left[ \bm{y}(t)=\bm{s}^l \right] | \left[ \bm{y}(t-1)=\bm{s}^k \right] \right) \\
  &\hphantom{AAAA}= \prod_{j=1}^n P\left(\left[ y_j(t) = s_j^l \right] | \left[ \bm{y}(t-1)=\bm{s}^k \right]\right) \\
  &\hphantom{AAAA}= \prod_{j=1}^n (a_j\bm{s}^k \text{ if } s_j^l=1 \text{ and } 1-a_j\bm{s}^k \text{ if } s_j^l = 1) \\
  &\hphantom{AAAA}= \prod_j^n (1-s_j^l) + (-1)^{s_j^l+1}a_j\bm{s}^k \text{.}
\end{align*}
Then, at $t$ for all $l$, the probability of the network being in any state is given by

\begin{align*}
  &P(\bm{y}(t)=\bm{s}^l) = \sum_{k=1}^n P(\left[\bm{y}(t-1)=\bm{s}^k\right])\\
  &\hphantom{P(\bm{y}(t)=\bm{s}^l) = \sum_{k=1}^n}
  P(\left[\bm{y}(t)=\bm{s}^l \right] | \left[\bm{y}(t-1)=\bm{s}^k\right]) \text{.}
\end{align*}

\subsection*{Characterizing distributions of cascade duration}
We characterized the distributions of cascade duration using a truncated power law. We used maximum likelihood estimation to estimate the power law with exponential cutoff $P(x) \sim x^{-\alpha}e^{x/\tau}$ \cite{clauset2009,alstott2014}. The exponent $\tau$ describes the value of $x$ at which the exponential cuts off the tail of the power law duration. To avoid overgeneralizing the extent of the power law, we bound $\tau$ by the maximum duration of $x$ and indicate this bound value as $\tau' = \min(\tau,\max(x))$.

\subsection*{Mutual information calculation to probe stimulus recovery}
To measure the capacity of a network to transfer information during a cascade, we calculated the mutual information $I(X;Y)$, which quantifies the amount of information, in bits, that one random variable $X$ reveals about another random variable $Y$. Here, the two random variables of interest are the initial stimulus patterns $\bm{y}(0) = \bm{s} \in S$ and the measured network states $\bm{y}(t) \in Y_t$ at each time step $t$. With mutual information, we measure the amount of information that the network states $Y_t$ reveal about which stimulus pattern $\bm{y}(0) \in \bm{s}_i$ or $S'$ was presented. To obtain a reasonable set of stimulus patterns, we generated the set $S' = \{\bm{s}_j | j \neq i\}$ of $\vert S' \vert = n-1$ unique stimulus patterns. Both $P(\bm{y}(0) = \bm{s}_i)$ and $P(\bm{y}(0) \in S')$ are equally probable at 0.5 such that $H(\bm{y}(0)) \approx 1$ bit for all patterns $i$. We generated the stimulus patterns such that they evenly spanned the range of average controllability values (see earlier Methods section on ``Stimulus pattern generation''). All mutual information calculations were run using the MIToolbox (v3.0.1) for MATLAB (https://github.com/Craigacp/MIToolbox).

In the analysis of the relationship between the average cascade duration and the mutual information, we quantify the decay in mutual information over time. We also calculate the correlation between the decay rate of the mutual information and the predicted mean cascade duration. For this latter calculation, first we perform a linear regression of the decay in mutual information with respect to time. Then, we calculate the Pearson correlation coefficient between the slope of the linear regression and the mean cascade duration.

\subsection*{Estimation by linear dynamical systems}
We prove by induction that linear dynamics estimates average behavior of the stochastic model, i.e., $\mathbb{E}[y_j(t)]=x_j(t)$, given the same initial conditions $\bm{y}(0)=\bm{x}(0)$. At $t=0$, both $y_j(0)$ and $x_j(0)$ are set as the stimulus pattern, and so, $\mathbb{E}[y_j(0)]=x_j(0)$. Now, assume $\mathbb{E}[y_j(t-1)]=x_j(t-1)$, and see that $x_j(t) = a_j^T \bm{x}(t-1) = a_j^T\mathbb{E}[\bm{y}(t-1)] = \mathbb{E}[a_j^T\bm{y}(t-1)] = \mathbb{E}[y_j(t)]$ and thus, $\mathbb{E}[y_j(t)]=x_j(t)$. To demonstrate this relation numerically, we take the average cascades that begin with the same initial state by taking the mean of $y_i^k(t)$ for all cascades $k$ at each time step $t$. All cascades start with the same initial condition $\bm{y}(0)$.

\subsection*{Eigenvalue analysis}
In our analysis of networks, we decompose the weight matrix $A$ into eigenvalues and eigenvectors. Such an eigendecomposition is formalized as

\begin{align}
A=PDP^{-1} \text{,}
\end{align}
where $P$ is a matrix of eigenvectors as columns and $D$ is a diagonal matrix of corresponding eigenvalues. We calculate the absolute value of the eigenvalue with the largest absolute value as the dominant eigenvalue $\lambda_1$.

When the row sum $\sum_j a_{ij}$ is greater than 1, the linear dynamical system does not equal the expected value of the stochastic model. However, the eigenvalue analyses can still be useful in describing average stochastic behavior. In particular, when $\lambda_1>1$, the state $\bm{x}(t)$ of the linear dynamical system can explode exponentially. While the state $\bm{y}(t)$ of a stochastic model with the same parameters does not similarly explode exponentially, it is bound by 1 for each neuron and reaches a fixed point at $\bm{1}$. In this case, the states of both models, $ \bm{x}(t) $ and $ \bm{y}(t) $, cannot reach quiescence at $ \bm{0} $ and thus have infinite cascade duration.

\subsection*{Network control theory and controllability statistics}
Network control theory is a formulation of control theory for networks of interacting components. This formulation typically consists of a set of $n$ component nodes $\mathcal V = \{1,\dotsm,n\}$, where the vector $\bm{x}(t) \in \real^n$ represents the state of node activities at time $t \geq 0$. These nodes are connected by a set of edges $\mathcal E \subseteq \mathcal V \times \mathcal V$, where the adjacency matrix $A \in \real^{n \times n}$ has elements $a_{ij}$ as the strength of the connection from node $j$ to node $i$. Here, \textit{control} typically refers to a set of $k$ inputs $\bm{u}(t) \in \real^k$ at time $t \geq 0$ that drive the evolution of system states according to $B\in\real^{n \times k}$. In linear control theory, the system states evolve as

\begin{align*}
\bm{x}(t+1) = A\bm{x}(t) + B\bm{u}(t).
\end{align*}

\subsection*{Finite average controllability}
Motivated by a desire to understand how network architecture affects its control properties, recent work iterates network-based metrics for control of such linear systems \cite{pasqualetti2014controllability}. Particularly germane to our discussion of cascade duration is \textit{average controllability} \cite{kailath1980,gu2015}, defined as the $H_2$ norm of the system's infinite average controllability given by

\begin{align*}
\textup{Trace}[W_K] =
\textup{Trace}\left[\sum_{\tau=0}^{\infty} A^\tau BB^T A^{T\tau}\right].
\end{align*}
Here, we set $B$ as a binary column vector where vector elements corresponding to the nodes of interest are set to 1 and the remaining vector elements are set to 0; this formulation represents an impulse of magnitude 1 to the nodes of interest. The finite average controllability (FAC) is similarly defined by taking the sum to some finite positive integer $F$ instead of infinity, and represents the norm of the system's impulse response over $F$ time steps. Because cascades are expected to last for a finite number of time steps, we use $F=100$ in the main text, and in the supplement we show that larger and smaller values of $\mathrm{F}$ produce similar results.

\subsection*{Modal controllability}
Another network-based control metric we use here is \textit{modal controllability} \cite{pasqualetti2014controllability, gu2015}. While modal controllability was originally formulated for symmetric matrices, here we extend the definition to include asymmetric matrices. To do this, we take the absolute value of both the eigenvalues and the eigenvector components, which can be complex numbers in an asymmetric matrix. Thus, we define the version of modal controllability of node $i$ for asymmetric matrices as

\begin{align*}
\phi_i = \sum^n_{j=1}(1-|\lambda_j|^2)|v_{ij}|^2 \text{.}
\end{align*}

\subsection*{Finite average controllability of initial states}
To predict the duration of a cascade, we can calculate the finite average controllability of an initial state $\bm{y}(0)$ defined as the finite average controllability averaged over the nodes that are active in the initial state,

\begin{align*}
\text{FAC}(\bm{y}(0)) = \frac{1}{|\bm{y}(0)|} \sum_{i \in \{i|y_i(0)=1\}} \text{FAC}_i \text{.}
\end{align*}
In the same way, we also calculate the finite average controllability in empirical cascades in the first $\{1...T\}$ time bins, averaging over the active neurons in those bins.

\section*{Code and data availability}

All code for simulations and analysis is publicly available at https://github.com/harangju/cascades. All data that support the findings of this study are available in the Open Science Framework with the identifier doi:10.17605/OSF.IO/TW69H.

\showmatmethods{} 

\acknow{H.J., J.Z.K., and D.S.B. acknowledge support from the John D. and Catherine T. MacArthur Foundation, the Alfred P. Sloan Foundation, the ISI Foundation, the Paul Allen Foundation, the Army Research Laboratory (W911NF-10-2-0022), the Army Research Office (Bassett-W911NF-14-1-0679, Grafton-W911NF-16-1-0474, DCIST- W911NF-17-2-0181), the Office of Naval Research, the National Institute of Mental Health (2-R01-DC-009209-11, R01-MH112847, R01-MH107235, R21-M MH-106799), the National Institute of Child Health and Human Development (1R01-HD086888-01), National Institute of Neurological Disorders and Stroke (R01-NS099348), and the National Science Foundation (BCS-1441502, BCS-1430087, NSF PHY-1554488 and BCS-1631550). J.Z.K acknowledges support from the NIH T32-EB020087, PD: Felix W. Wehrli, and the National Science Foundation Graduate Research Fellowship No. DGE-1321851. We thank Erin Teich, Lia Papadopoulos, and Zhixin Lu for comments and suggestions on the paper. The content is solely the responsibility of the authors and does not necessarily represent the official views of any of the funding agencies.}

\showacknow{} 

\nolinenumbers

\section*{}

\bibliography{library}

\newcommand{\noop}[1]{}
\begin{thebibliography}{100}

\bibitem{adolphs2003}
Adolphs R (2003) Cognitive neuroscience of human social behaviour.
\newblock {\em Nature Reviews Neuroscience} 4:165 EP --.

\bibitem{watts1998}
Watts DJ, Strogatz SH (1998) Collective dynamics of `small-world'networks.
\newblock {\em Nature} 393:440 EP --.

\bibitem{honey2007}
Honey CJ, K{\"o}tter R, Breakspear M, Sporns O (2007) Network structure of
  cerebral cortex shapes functional connectivity on multiple time scales.
\newblock {\em Proceedings of the National Academy of Sciences}
  104(24):10240--10245.

\bibitem{hopfield1982}
Hopfield JJ (1982) Neural networks and physical systems with emergent
  collective computational abilities.
\newblock {\em Proceedings of the National Academy of Sciences}
  79(8):2554--2558.

\bibitem{benyishai1995}
Ben-Yishai R, Bar-Or RL, Sompolinsky H (1995) Theory of orientation tuning in
  visual cortex.
\newblock {\em Proceedings of the National Academy of Sciences}
  92(9):3844--3848.

\bibitem{wang2002}
Wang XJ (2002) Probabilistic decision making by slow reverberation in cortical
  circuits.
\newblock {\em Neuron} 36(5):955--968.

\bibitem{haldeman2005}
Haldeman C, Beggs JM (2005) Critical branching captures activity in living
  neural networks and maximizes the number of metastable states.
\newblock {\em Phys. Rev. Lett.} 94(5):058101.

\bibitem{beggs2003}
Beggs JM, Plenz D (2003) Neuronal avalanches in neocortical circuits.
\newblock {\em Journal of Neuroscience} 23(35):11167--11177.

\bibitem{beggs2004}
Beggs JM, Plenz D (2004) Neuronal avalanches are diverse and precise activity
  patterns that are stable for many hours in cortical slice cultures.
\newblock {\em Journal of Neuroscience} 24(22):5216--5229.

\bibitem{gireesh2008}
Gireesh ED, Plenz D (2008) Neuronal avalanches organize as nested theta- and
  beta/gamma-oscillations during development of cortical layer 2/3.
\newblock {\em Proceedings of the National Academy of Sciences}
  105(21):7576--7581.

\bibitem{petermann2009}
Petermann T, et~al. (2009) Spontaneous cortical activity in awake monkeys
  composed of neuronal avalanches.
\newblock {\em Proceedings of the National Academy of Sciences}
  106(37):15921--15926.

\bibitem{hahn2010}
Hahn G, et~al. (2010) Neuronal avalanches in spontaneous activity in vivo.
\newblock {\em Journal of Neurophysiology} 104(6):3312--3322.
\newblock PMID: 20631221.

\bibitem{shriki2013}
Shriki O, et~al. (2013) Neuronal avalanches in the resting meg of the human
  brain.
\newblock {\em Journal of Neuroscience} 33(16):7079--7090.

\bibitem{bellay2015}
Bellay T, Klaus A, Seshadri S, Plenz D (2015) Irregular spiking of pyramidal
  neurons organizes as scale-invariant neuronal avalanches in the awake state.
\newblock {\em eLife} 4:e07224.

\bibitem{poncealvarez2018}
Ponce-Alvarez A, Jouary A, Privat M, Deco G, Sumbre G (2018) Whole-brain
  neuronal activity displays crackling noise dynamics.
\newblock {\em Neuron}.

\bibitem{shew2015}
Shew WL, et~al. (2015) Adaptation to sensory input tunes visual cortex to
  criticality.
\newblock {\em Nature Physics} 11:659 EP --.

\bibitem{shew2011}
Shew WL, Yang H, Yu S, Roy R, Plenz D (2011) Information capacity and
  transmission are maximized in balanced cortical networks with neuronal
  avalanches.
\newblock {\em Journal of Neuroscience} 31(1):55--63.

\bibitem{bertschinger2004}
Bertschinger N, Natschl{\"a}ger T (2004) Real-time computation at the edge of
  chaos in recurrent neural networks.
\newblock {\em Neural Computation} 16(7):1413--1436.

\bibitem{kinouchi2006}
Kinouchi O, Copelli M (2006) Optimal dynamical range of excitable networks at
  criticality.
\newblock {\em Nature Physics} 2:348 EP --.

\bibitem{shew2009}
Shew WL, Yang H, Petermann T, Roy R, Plenz D (2009) Neuronal avalanches imply
  maximum dynamic range in cortical networks at criticality.
\newblock {\em Journal of Neuroscience} 29(49):15595--15600.

\bibitem{larremore2011b}
Larremore DB, Shew WL, Ott E, Restrepo JG (2011) Effects of network topology,
  transmission delays, and refractoriness on the response of coupled excitable
  systems to a stochastic stimulus.
\newblock {\em Chaos (Woodbury, N.Y.)} 21(2):025117--025117.

\bibitem{wang2006}
Wang Y, et~al. (2006) Heterogeneity in the pyramidal network of the medial
  prefrontal cortex.
\newblock {\em Nature Neuroscience} 9:534 EP --.

\bibitem{lefort2009}
Lefort S, Tomm C, Sarria JCF, Petersen CC (2009) The excitatory neuronal
  network of the c2 barrel column in mouse primary somatosensory cortex.
\newblock {\em Neuron} 61(2):301 -- 316.

\bibitem{ko2011}
Ko H, et~al. (2011) Functional specificity of local synaptic connections in
  neocortical networks.
\newblock {\em Nature} 473:87 EP --.

\bibitem{brunel2016}
Brunel N (2016) Is cortical connectivity optimized for storing information?
\newblock {\em Nature Neuroscience} 19:749 EP --.

\bibitem{markram1997}
Markram H (1997) A network of tufted layer 5 pyramidal neurons.
\newblock {\em Cerebral Cortex} 7(6):523--533.

\bibitem{song2005}
Song S, Sj{\"o}str{\"o}m PJ, Reigl M, Nelson S, Chklovskii DB (2005) Highly
  nonrandom features of synaptic connectivity in local cortical circuits.
\newblock {\em PLOS Biology} 3(3).

\bibitem{perin2011}
Perin R, Berger TK, Markram H (2011) A synaptic organizing principle for
  cortical neuronal groups.
\newblock {\em Proceedings of the National Academy of Sciences}
  108(13):5419--5424.

\bibitem{shimono2014}
Shimono M, Beggs JM (2014) {Functional Clusters, Hubs, and Communities in the
  Cortical Microconnectome}.
\newblock {\em Cerebral Cortex} 25(10):3743--3757.

\bibitem{nigam2016}
Nigam S, et~al. (2016) Rich-club organization in effective connectivity among
  cortical neurons.
\newblock {\em Journal of Neuroscience} 36(3):670--684.

\bibitem{faber2019}
Faber SP, Timme NM, Beggs JM, Newman EL (2018) Computation is concentrated in
  rich clubs of local cortical networks.
\newblock {\em Network Neuroscience} 3(2):384--404.

\bibitem{touboul2010}
Touboul J, Destexhe A (2010) Can power-law scaling and neuronal avalanches
  arise from stochastic dynamics?
\newblock {\em PLOS ONE} 5(2):1--14.

\bibitem{friedman2012}
Friedman N, et~al. (2012) Universal critical dynamics in high resolution
  neuronal avalanche data.
\newblock {\em Phys. Rev. Lett.} 108(20):208102.

\bibitem{priesemann2014}
Priesemann V, et~al. (2014) Spike avalanches in vivo suggest a driven, slightly
  subcritical brain state.
\newblock {\em Frontiers in Systems Neuroscience} 8:108.

\bibitem{touboul2017}
Touboul J, Destexhe A (2017) Power-law statistics and universal scaling in the
  absence of criticality.
\newblock {\em Phys. Rev. E} 95(1):012413.

\bibitem{goldmanrakic1995}
Goldman-Rakic P (1995) Cellular basis of working memory.
\newblock {\em Neuron} 14(3):477 -- 485.

\bibitem{durstewitz2000}
Durstewitz D, Seamans JK, Sejnowski TJ (2000) Neurocomputational models of
  working memory.
\newblock {\em Nature Neuroscience} 3:1184 EP --.

\bibitem{eriksson2015}
Eriksson J, Vogel EK, Lansner A, Bergstr{\"o}m F, Nyberg L (2015)
  Neurocognitive architecture of working memory.
\newblock {\em Neuron} 88(1):33 -- 46.

\bibitem{mcculloch1943}
McCulloch WS, Pitts W (1990) A logical calculus of the ideas immanent in
  nervous activity. 1943.
\newblock {\em Bulletin of mathematical biology} 52 1-2:99--115; discussion
  73--97.

\bibitem{garlaschelli2009}
Garlaschelli D (2009) The weighted random graph model.
\newblock {\em New Journal of Physics} 11(7):073005.

\bibitem{seung1996}
Seung HS (1996) How the brain keeps the eyes still.
\newblock {\em Proceedings of the National Academy of Sciences} 93(23):13339.

\bibitem{larremore2011a}
Larremore DB, Shew WL, Restrepo JG (2011) Predicting criticality and dynamic
  range in complex networks: Effects of topology.
\newblock {\em Phys. Rev. Lett.} 106(5):058101.

\bibitem{clauset2009}
Clauset A, Shalizi C, Newman M (2009) Power-law distributions in empirical
  data.
\newblock {\em SIAM Review} 51(4):661--703.

\bibitem{alstott2014}
Alstott J, Bullmore E, Plenz D (2014) powerlaw: A python package for analysis
  of heavy-tailed distributions.
\newblock {\em PLOS ONE} 9(1):1--11.

\bibitem{bak1987}
Bak P, Tang C, Wiesenfeld K (1987) Self-organized criticality: An explanation
  of the 1/f noise.
\newblock {\em Physical Review Letters} 59(4):381--384.

\bibitem{ito2016}
Ito S, et~al. (2016) Spontaneous spiking activity of hundreds of neurons in
  mouse somatosensory cortex slice cultures recorded using a dense 512
  electrode array.
\newblock {\em CRCNS.org}.

\bibitem{neumaier2001}
Neumaier A, Schneider T (2001) Estimation of parameters and eigenmodes of
  multivariate autoregressive models.
\newblock {\em ACM Trans. Math. Softw.} 27(1):27--57.

\bibitem{schneider2001}
Schneider T, Neumaier A (2001) Algorithm 808: Arfit\&mdash;a matlab package for
  the estimation of parameters and eigenmodes of multivariate autoregressive
  models.
\newblock {\em ACM Trans. Math. Softw.} 27(1):58--65.

\bibitem{bassett2013}
Bassett DS, et~al. (2013) Robust detection of dynamic community structure in
  networks.
\newblock {\em Chaos} 23(1):013142.

\bibitem{vankessenich2016}
Michiels~van Kessenich L, de~Arcangelis L, Herrmann HJ (2016) Synaptic
  plasticity and neuronal refractory time cause scaling behaviour of neuronal
  avalanches.
\newblock {\em Scientific Reports} 6:32071 EP --.

\bibitem{connors1990}
Connors BW, Gutnick MJ (1990) Intrinsic firing patterns of diverse neocortical
  neurons.
\newblock {\em Trends in Neurosciences} 13(3):99--104.

\bibitem{poil2012}
Poil SS, Hardstone R, Mansvelder HD, Linkenkaer-Hansen K (2012) Critical-state
  dynamics of avalanches and oscillations jointly emerge from balanced
  excitation/inhibition in neuronal networks.
\newblock {\em Journal of Neuroscience} 32(29):9817--9823.

\bibitem{lombardi2014}
Lombardi F, Herrmann HJ, Plenz D, De~Arcangelis L (2014) On the temporal
  organization of neuronal avalanches.
\newblock {\em Frontiers in Systems Neuroscience} 8:204.

\bibitem{pasqualetti2014controllability}
Pasqualetti F, Zampieri S, Bullo F (2014) Controllability metrics, limitations
  and algorithms for complex networks.
\newblock {\em IEEE Transactions on Control of Network Systems} 1(1):40--52.

\bibitem{gu2015}
Gu S, et~al. (2015) Controllability of structural brain networks.
\newblock {\em Nature Communications} 6:8414 EP --.

\bibitem{tang2017developmental}
Tang E, et~al. (2017) Developmental increases in white matter network
  controllability support a growing diversity of brain dynamics.
\newblock {\em Nat Commun} 8(1):1252.

\bibitem{wang2013}
Wang Z, et~al. (2013) The relationship of anatomical and functional
  connectivity to resting-state connectivity in primate somatosensory cortex.
\newblock {\em Neuron} 78(6):1116 -- 1126.

\bibitem{lombard2012}
Lombardi F, Herrmann HJ, Perrone-Capano C, Plenz D, de~Arcangelis L (2012)
  Balance between excitation and inhibition controls the temporal organization
  of neuronal avalanches.
\newblock {\em Phys. Rev. Lett.} 108(22):228703.

\bibitem{liu2011}
Liu YY, Slotine JJ, Barab{\'a}si AL (2011) Controllability of complex networks.
\newblock {\em Nature} 473:167 EP --.

\bibitem{yan2017}
Yan G, et~al. (2017) Network control principles predict neuron function in the
  {Caenorhabditis} elegans connectome.
\newblock {\em Nature} 550:519 EP --.

\bibitem{wiles2017autaptic}
Wiles L, et~al. (2017) Autaptic connections shift network excitability and
  bursting.
\newblock {\em Sci Rep} 7:44006.

\bibitem{towlson2018celegans}
Towlson EK, et~al. (2018) Caenorhabditis elegans and the network control
  framework-{FAQs}.
\newblock {\em Philos Trans R Soc Lond B Biol Sci} 373:1758.

\bibitem{cornblath2018sex}
Cornblath EJ, et~al. (2018) Sex differences in network controllability as a
  predictor of executive function in youth.
\newblock {\em Neuroimage} 188:122--134.

\bibitem{jeganathan2018fronto}
Jeganathan J, et~al. (2018) Fronto-limbic dysconnectivity leads to impaired
  brain network controllability in young people with bipolar disorder and those
  at high genetic risk.
\newblock {\em Neuroimage Clin} 19:71--81.

\bibitem{tang2018control}
Tang E, Bassett DS (2018) Control of dynamics in brain networks.
\newblock {\em Rev. Mod. Phys.} 90:031003.

\bibitem{chen1998}
Chen CT (1998) {\em Linear System Theory and Design}.
\newblock (Oxford University Press, Inc., New York, NY, USA), 3rd edition.

\bibitem{taylor2015optimal}
Taylor PN, et~al. (2015) Optimal control based seizure abatement using patient
  derived connectivity.
\newblock {\em Front Neurosci} 9:202.

\bibitem{betzel2016optimally}
Betzel RF, Gu S, Medaglia JD, Pasqualetti F, Bassett DS (2016) Optimally
  controlling the human connectome: the role of network topology.
\newblock {\em Sci Rep} 6:30770.

\bibitem{gu2017optimal}
Gu S, et~al. (2017) Optimal trajectories of brain state transitions.
\newblock {\em Neuroimage} 148:305--317.

\bibitem{medaglia2018network}
Medaglia JD, et~al. (2018) Network controllability in the inferior frontal
  gyrus relates to controlled language variability and susceptibility to {TMS}.
\newblock {\em J Neurosci} 38(28):6399--6410.

\bibitem{muldoon2016stimulation}
Muldoon SF, et~al. (2016) Stimulation-based control of dynamic brain networks.
\newblock {\em PLoS Comput Biol} 12(9):e1005076.

\bibitem{stiso2018white}
Stiso J, et~al. (2018) White matter network architecture guides direct
  electrical stimulation through optimal state transitions.
\newblock {\em arXiv} 1805:01260.

\bibitem{khambhati2018predictive}
Khambhati AN, et~al. (ahead of print) Predictive control of
  electrophysiological network architecture using direct, single-node
  neurostimulation in humans.
\newblock {\em Network Neuroscience}
  https://www.mitpressjournals.org/doi/abs/10.1162/netn\_a\_00089.

\bibitem{alon2007network}
Alon U (2007) Network motifs: theory and experimental approaches.
\newblock {\em Nat Rev Genet} 8(6):450--461.

\bibitem{shenorr2002network}
Shen-Orr SS, Milo R, Mangan S, Alon U (2002) Network motifs in the
  transcriptional regulation network of {E}scherichia coli.
\newblock {\em Nat Genet} 31(1):64--68.

\bibitem{yeger2004network}
Yeger-Lotem E, et~al. (2004) Network motifs in integrated cellular networks of
  transcription-regulation and protein-protein interaction.
\newblock {\em Proc Natl Acad Sci U S A} 101(16):5934--5939.

\bibitem{hart2012design}
Hart Y, Antebi YE, Mayo AE, Friedman N, Alon U (2012) Design principles of cell
  circuits with paradoxical components.
\newblock {\em Proc Natl Acad Sci U S A} 109(21):8346--8351.

\bibitem{kashtan2005spontaneous}
Kashtan N, Alon U (2005) Spontaneous evolution of modularity and network
  motifs.
\newblock {\em Proc Natl Acad Sci U S A} 102(39):13773--13778.

\bibitem{sporns2004motifs}
Sporns O, Kotter R (2004) Motifs in brain networks.
\newblock {\em PLoS Biol} 2(11):e369.

\bibitem{sizemore2018cliques}
Sizemore AE, et~al. (2018) Cliques and cavities in the human connectome.
\newblock {\em J Comput Neurosci} 44(1):115--145.

\bibitem{giusti2016twos}
Giusti C, Ghrist R, Bassett DS (2016) Two's company, three (or more) is a
  simplex: {A}lgebraic-topological tools for understanding higher-order
  structure in neural data.
\newblock {\em J Comput Neurosci} 41(1):1--14.

\bibitem{sizemore2018importance}
Sizemore AE, Phillips-Cremins JE, Ghrist R, Bassett DS (2018) The importance of
  the whole: Topological data analysis for the network neuroscientist.
\newblock {\em Network Neuroscience} Epub Ahead of Print.

\bibitem{reimann2017cliques}
Reimann MW, et~al. (2017) Cliques of neurons bound into cavities provide a
  missing link between structure and function.
\newblock {\em Front Comput Neurosci} 11:48.

\bibitem{shannon1948}
Shannon CE (1948) A mathematical theory of communication.
\newblock {\em Bell System Technical Journal} 27(3):379--423.

\bibitem{timme2016}
Timme NM, et~al. (2016) High-degree neurons feed cortical computations.
\newblock {\em PLOS Computational Biology} 12(5):1--31.

\bibitem{kim2017}
Kim JZ, et~al. (2018) Role of graph architecture in controlling dynamical
  networks with applications to neural systems.
\newblock {\em Nature Physics} 14:91--98.

\bibitem{hodgkin1952}
HODGKIN AL, HUXLEY AF (1952) A quantitative description of membrane current and
  its application to conduction and excitation in nerve.
\newblock {\em The Journal of physiology} 117(4):500--544.

\bibitem{motter2015}
Motter AE (2015) Networkcontrology.
\newblock {\em Chaos (Woodbury, N.Y.)} 25(9):097621; 097621--097621.

\bibitem{hebb1949}
Hebb D (1949) {\em The Organization of Behavior: a Neuropsychological Theory}.
\newblock (Oxford, England: Wiley).

\bibitem{song2000}
Song S, Miller KD, Abbott LF (2000) Competitive hebbian learning through
  spike-timing-dependent synaptic plasticity.
\newblock {\em Nature Neuroscience} 3:919 EP --.

\bibitem{wuyan2018}
Wu-Yan E, et~al. (2018) Benchmarking measures of network controllability on
  canonical graph models.
\newblock {\em Journal of Nonlinear Science}.

\bibitem{schwarz1978}
Schwarz G (1978) Estimating the dimension of a model.
\newblock {\em The Annals of Statistics} 6(2):461--464.

\bibitem{brunel2004}
Brunel N, Hakim V, Isope P, Nadal JP, Barbour B (2004) Optimal information
  storage and the distribution of synaptic weights: Perceptron versus purkinje
  cell.
\newblock {\em Neuron} 43(5):745--757.

\bibitem{iyer2013}
Iyer R, Menon V, Buice M, Koch C, Mihalas S (2013) The influence of synaptic
  weight distribution on neuronal population dynamics.
\newblock {\em PLoS computational biology} 9(10):e1003248; e1003248--e1003248.

\bibitem{pehlevan2017}
Pehlevan C, Sengupta A (2017) Resource-efficient perceptron has sparse synaptic
  weight distribution.
\newblock {\em 2017 25th Signal Processing and Communications Applications
  Conference (SIU)} pp. 1--4.

\bibitem{vanrossum2000}
van Rossum MCW, Bi GQ, Turrigiano GG (2000) Stable hebbian learning from spike
  timing-dependent plasticity.
\newblock {\em The Journal of Neuroscience} 20(23):8812.

\bibitem{markram1997a}
Markram H, L{\"u}bke J, Frotscher M, Roth A, Sakmann B (1997) Physiology and
  anatomy of synaptic connections between thick tufted pyramidal neurones in
  the developing rat neocortex.
\newblock {\em The Journal of physiology} 500 ( Pt 2)(Pt 2):409--440.

\bibitem{feldmeyer1999}
Feldmeyer D, Egger V, L{\"u}bke J, Sakmann B (1999) Reliable synaptic
  connections between pairs of excitatory layer 4 neurones within a single
  `barrel'of developing rat somatosensory cortex.
\newblock {\em The Journal of Physiology} 521(1):169--190.

\bibitem{chen2010}
Chen W, Hobbs JP, Tang A, Beggs JM (2010) A few strong connections: optimizing
  information retention in neuronal avalanches.
\newblock {\em BMC neuroscience} 11:3--3.

\bibitem{kailath1980}
Kailath T (1980) {\em Linear Systems}.
\newblock (Prentice-Hall).

\end{thebibliography}

\end{document}